\documentclass{amia}
\usepackage[labelfont=bf]{caption}
\usepackage{color}

\usepackage[super,comma]{natbib}
\usepackage{times}
\usepackage{soul}
\usepackage{url}
\usepackage[hidelinks]{hyperref}
\usepackage[utf8]{inputenc}
\usepackage{amsmath}
\usepackage{amsthm}
\usepackage{booktabs}
\usepackage{algorithm}
\usepackage{algorithmic}
\usepackage{multirow}
\usepackage{subcaption}
\usepackage{makecell}
\usepackage{amsfonts}
\usepackage{amssymb}
\usepackage{graphicx}
\usepackage{wrapfig}

\begin{document}

\setlength{\bibsep}{0.1pt}

\title{DiffusionCT: Latent Diffusion Model for CT Image Standardization}

\author{
Md Selim$^{1,3}$, 
Jie Zhang, PhD$^2$, 
Michael A. Brooks, MD$^{2}$, 
Ge Wang, PhD$^5$, 
Jin Chen, PhD$^{1,3,4}$}

\institutes{
$^1$Department of Computer Science 
$^2$Department of Radiology 
$^3$Institute for Biomedical Informatics 
$^4$Department of Internal Medicine,  University of Kentucky, Lexington, KY 
$^5$Biomedical Imaging Center, Rensselaer Polytechnic Institute, Troy, NY 
}

\maketitle

\noindent{\bf Abstract}
\textit{
Computed tomography (CT) is one of the modalities for effective lung cancer screening, diagnosis, treatment, and prognosis. The features extracted from CT images are now used to quantify spatial and temporal variations in tumors. However, CT images obtained from various scanners with customized acquisition protocols may introduce considerable variations in texture features, even for the same patient. This presents a fundamental challenge to downstream studies that require consistent and reliable feature analysis.
%
%
Existing CT image harmonization models rely on GAN-based supervised or semi-supervised learning, with limited performance. This work addresses the issue of CT image harmonization using a new diffusion-based model, named DiffusionCT, to standardize CT images acquired from different vendors and protocols. DiffusionCT operates in the latent space by mapping a latent non-standard distribution into a standard one. 
DiffusionCT incorporates an Unet-based encoder-decoder, augmented by a diffusion model integrated into the bottleneck part. The model is designed in two training phases. The encoder-decoder is first trained, without embedding the diffusion model, to learn the latent representation of the input data. The latent diffusion model is then trained in the next training phase while fixing the encoder-decoder. 
Finally, the decoder synthesizes a standardized image with the transformed latent representation.
The experimental results demonstrate a significant improvement in the performance of the standardization task using DiffusionCT. 
}

\section{Introduction}\label{sec:intro}
Lung cancer is the leading cause of cancer death and one of the most common cancers among both men and women in the United States~\cite{collins2017letter}. The overall 5-year survival rate for non-small cell lung cancer is approximately 19\%. Computed tomography (CT) imaging is a critical tool for early lung cancer diagnosis and tumor phenotype characterization for improved treatment and prognosis~\cite{de2014benefits,ravanelli2013texture}. Deep texture features extracted from CT images using advanced deep learning algorithms can quantify spatial and temporal variations in tumor architecture and function, allowing for the determination of intra-tumor evolution~\cite{ardila2019end,song2017using}. However, CT images are often acquired using scanners from different vendors with customized acquisition standards, resulting in considerable variations in texture features, even for the same patient. This variability poses a significant challenge for large-scale studies across different sites~\cite{lu2020deep}. The lack of standardized radiomics is a significant obstacle in downstream clinical tasks.

\begin{figure}[!bt]
\centering
\includegraphics[width=.7\textwidth]{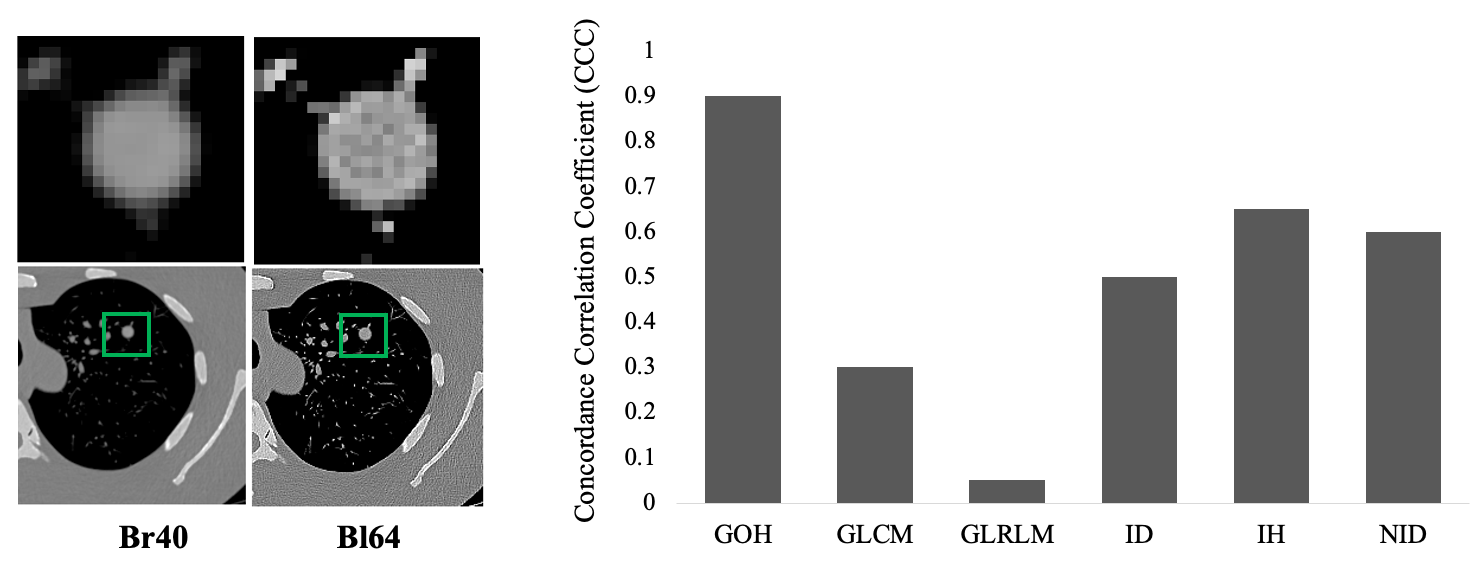}
\caption{{\bf Discrepancy of tumor image features caused by different imaging protocols.} The same lungman chest phantom was scanned using the same scanner. CT images were acquired using two different image reconstruction kernels accordingly, as indicated by the texts at the bottom of the images. In the images, a tumor is marked with green rectangles. The histogram on the right side showed the feature variance between these two tumors in terms of CCC. The observed differences in the tumor images may have significant implications on the promise of large-scale radiomic studies.  } \label{fig:senario} 
\end{figure}

Radiomic inconsistency in texture, shape, and intensity has been observed among images acquired with scanners from different vendors using non-standardized protocols~\cite{berenguer2018radiomics,hunter2013high}.  
%
The intra-scanner (different acquisition protocols of the same scanner) and cross-scanner (similar acquisition protocols of different scanners) variation remains a challenge to be solved. Figure~\ref{fig:senario} shows an example of the impact of non-standard CT imaging acquisition protocols on radiomic features. 
A Lungman chest phantom with three artificial tumors was scanned using CT scanners manufactured by Siemens. Images were reconstructed using Bl64 and Br40 reconstruction kernels. The tumors' visual appearance and radiomic features in these images differed significantly. 
%

Developing a universal standard for CT image acquisition has been suggested as a potential solution to the CT image radiomic feature discrepancy problem. However, implementing such a standard would require significant modifications to existing CT imaging protocols, and could potentially limit the range of applications for the modality~\cite{paul2012relationships,gierada2010effects}. In light of these limitations, alternative approaches are needed to address the issue of radiomic feature discrepancies in CT images.
%

Recent advancement has been made to address the CT image inconsistency problem. One solution is to develop a post-processing framework that can standardize and normalize existing CT images while preserving anatomic details~\cite{ours_aamp,cohen2012radiosity,selim2020stan,selim2021cross,selim2021radiomic}. Our research has shown that this approach allows for the extraction of reliable and consistent features from standardized images, facilitating accurate downstream analysis, and ultimately leading to improved diagnosis, treatment, and prognosis of lung cancer. Deep learning-based image standardization algorithms hold great promise in harmonizing CT images acquired with different parameters from the same scanner. 
%
It is essential to acknowledge that the existing solutions exhibit limitations in terms of image texture synthesis and structure maintenance performance. These constraints can adversely affect the accuracy of subsequent analyses, thereby impeding the development of dependable and consistent features that are crucial for enhancing lung cancer diagnosis, treatment, and prognosis.
Therefore, further research is needed to develop more advanced algorithms to address these challenges and augment the performance of CT image standardization. Progress in this domain can lead to significant improvements in the realm of medical imaging and contribute to the development of more effective strategies for combating lung cancer.

\begin{figure}[bt!]
\centering
\includegraphics[width=\textwidth]{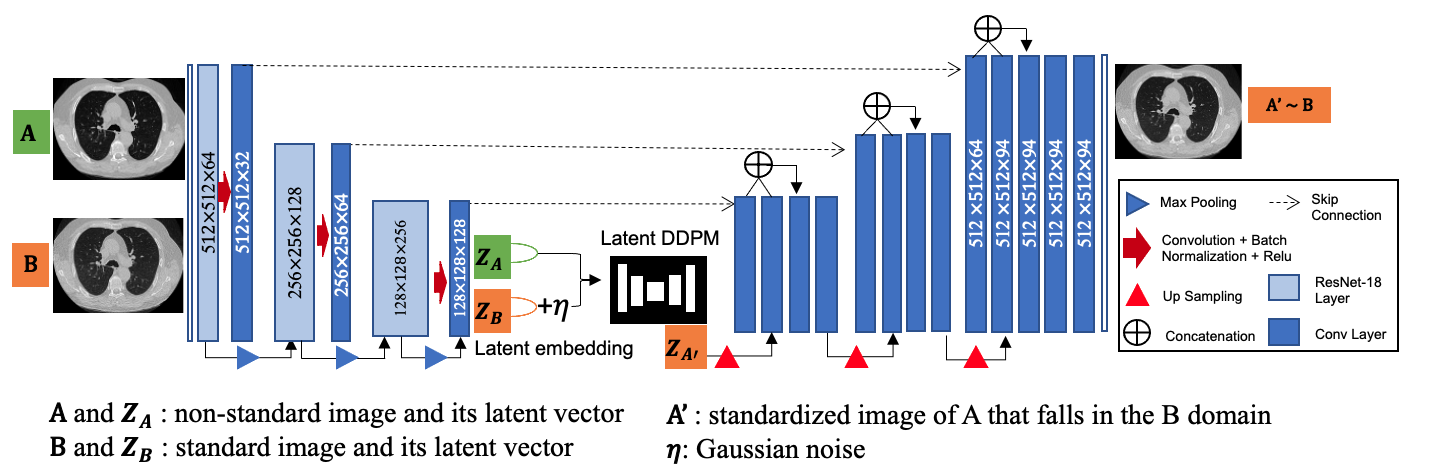}
\caption{Overview of a DDPM pipeline for intra-scanner standardization. Given an image pair (A, B) where A and B are non-standard and the corresponding standard images, the model aims to synthesize a new image $A'$ in domain $B$. The representation learning component learns encoded latent representations of CT images using a ResNet-18-based encoder-decoder structure. The target-specific latent-space mapping component is designed for standard image synthesis. It contains a DDPM model for latent space mapping. $Z_A$ is the latent vector of non-standard image A; $Z_B$ is the latent vector of standard image B; $Z_{A'}$ is the standardized latent vector of image A, and $\eta$ is Gaussian noise. } \label{fig:diff_model}
\end{figure}

Compared to the state-of-the-art generative adversarial networks (GAN) and variational auto-encoders (VAE) algorithms, DDPM~\cite{ho2020denoising} shows superior performance in image standardization. DDPM learns a Markov chain to gradually convert a simple distribution, such as isotropic Gaussian, into a target data distribution. It consists of two processes: (1) a fixed forward diffusion process that gradually adds noise to an image when sequentially sampling latent variables of the same dimensionality and (2) a learned reverse denoising diffusion process, where a neural network (such as U-Net) is trained to gradually denoise an image starting from a pure noise realization. DDPM and its variants have attracted a surge of attention since 2020, resulting in key advances in continuous data modeling, such as image generation~\cite{ho2020denoising}, super-resolution~\cite{saharia2022image}, and image-to-image translation~\cite{saharia2022palette}. More recently, conditional DDPM has shown remarkable performance in conditional image generation. In parallel, latent DDPM enables generating image embedding in a low-dimensional latent space. 
%

Leveraging the recent advances in the denoising diffusion probabilistic models (DDPM), the present study proposes a novel algorithm named DiffusionCT for CT image standardization ~\cite{ho2020denoising}. DiffusionCT comprises an encoder-decoder network and a latent conditional DDPM, and its network architecture is depicted in Fig~\ref{fig:diff_model}. 
The encoder-decoder network maps the input CT image to a low-dimensional latent representation. The DDPM then models the conditional probability distribution of the latent representation to synthesize a standard image. Notably, DiffusionCT preserves the original structure of the CT image while effectively standardizing texture.

The training of DiffusionCT consists of three phases. First,  an encoder-decoder network is trained with all training CT images, no matter whether they are standard or non-standard. This process aims to encode every input image into a 1-D latent vector effectively that can reconstruct the original image with minimal information loss. Second, a latent conditional DDPM is trained with image pairs, where one is a non-standard image, and the other is the corresponding standard image. This process enables the DDPM to model the conditional probability distribution of the latent representation, allowing it to synthesize standard images. Finally,  all the trained neural networks are integrated to standardize new images.

%

\section{Background} \label{sec_bcg}

\subsection{Radiomic Features} 
Radiology, as a branch of medicine, employs sophisticated non-invasive imaging technologies for the diagnosis and treatment of various diseases. Crucial to tumor characterization are the image features extracted from radiological images~\cite{yip2016applications}. 
Among these features, radiomic features provide insight into the cellular and genetic levels of phenotypic patterns hidden from the naked eyes~\cite{ yang2011quantifying, basu2011evolving, yip2016applications}. 
Radiomic features can be categorized into six classes: Gradient Oriented Histogram (GOH), Gray Level Co-occurrence Matrix (GLCM), Gray Level Run Length Matrix (GLRLM), Intensity Direct (ID), Intensity Histogram (IH), and Neighbor Intensity Difference (NID). 
Utilizing radiomic features offers considerable potential to capture tumor heterogeneity and detailed phenotypic information. 
However, the efficacy of radiomic studies, especially in the context of extensive cross-institutional collaborations, is significantly hindered by the lack of standardization in medical image acquisition practices~\cite{hunter2013high, berenguer2018radiomics}. 


\subsection{CT Image Standardization Approaches} 
In general, There are two types of CT image standardization approaches, each serving distinct purposes and contingent upon data availability. 
The first category, known as intra-scanner image standardization, necessitates the availability of paired image data~\cite{selim2020stan}. In this scenario, two images constructed from the same scan but employing different reconstruction kernels constitute an image pair, where the source image refers to the image constructed with the non-standard kernel (e.g., Siemens Br40), and the target image is constructed  using the standard kernel (e.g., Siemens Bl64). Given paired image data as the training data, a machine learning model is trained to convert source images to target images. 
The second category for CT image standardization encompasses models devised for cross-scanner image standardization, which eliminates the need for paired image data~\cite{selim2021cross}. In this setting, images are not required to be matched; rather, images acquired with different protocols are stored separately. 

Acquiring paired training data is straightforward, though it is predominantly confined to a single scanner. In large-scale radiomic studies, the need for standardization is more pronounced in cross-vendor scenarios, which cannot be accomplished by utilizing models from the first category. To address the issue of cross-vendor image standardization, models in the second category mitigate the requirement for paired images, albeit at the cost of reduced performance.

%
Liang et al~\cite{liang2019ganai} developed a CT image standardization model, denoted as GANai, based on conditional Generative Adversarial Network (cGAN)~\cite{pix2pix}. A new alternative training strategy was designed to effectively learn the data distribution. GANai achieved better performance in comparison to cGAN and the traditional histogram matching approach~\cite{gonzalez2012digital}. However, GANai primarily focuses on the less challenging task of image patch synthesis  rather than addressing the entire DICOM image synthesis problem.

Selim et al~\cite{selim2020stan} introduced another cGAN-based CT image standardization model, denoted as STAN-CT. In STAN-CT, a complete pipeline for systematic CT image standardization was constructed. Also, a new loss function was devised to account for two constraints, i.e., latent space loss and feature space loss. The latent space loss is adopted for the generator to establish a one-to-one mapping between  standard and synthesized images. The feature space loss is utilized by the discriminator to critique the texture features of the standard and the synthesized images. Nevertheless, STAN-CT was limited by the limited availability of training data and was evaluated at the image patch level on a limited number of texture features, utilizing only a single evaluation criterion. 

RadiomicGAN, another GAN-based model, incorporates a transfer learning approach to address the data limitation issue~\cite{selim2021radiomic}. The model is designed using a pre-trained VGG network. A novel training technique called window training is implemented to reconcile the pixel intensity disparity between the natural image domain and the CT imaging domain.  Experimental results indicated that RadiomicGAN outperformed both STAN-CT and GANai.

For cross-scanner image standardization, a model termed CVH-CT was developed~\cite{selim2021cross}. CVH-CT aims to standardize images between scanners from different manufacturers, such as Siemens and GE. The generator of CVH-CT employs a self-attention mechanism for learning scanner-related information. A VGG feature-based domain loss is utilized to extract texture properties from unpaired image data, enabling the learning of scanner-based texture distributions. Experimental results show that, in comparison to CycleGAN~\cite{CycleGAN2017}, CVH-CT enhanced feature discrepancy in the synthesized images, but its performance is not significantly improved when compared with models trained within the intra-scanner domain. 

UDA-CT, a recently developed deep learning model for CT image standardization, demonstrates a departure from previous methods by incorporating both paired and unpaired images, rendering it more flexible and robust~\cite{selim2022udact}. UDA-CT effectively learns a mapping from all non-standard distributions to the standard distribution, thereby enhancing the modeling of the global distribution of all non-standard images. Notably, UDA-CT demonstrates compatible performance in both within-scanner and cross-scanner settings. 


The development of standardization models for CT images has provided a solid foundation for generating stable radiomic features in large-scale studies. However, recent advances in image synthesis using diffusion models have opened up new opportunities for investigating the CT image standardization problem. These models offer a powerful approach for generating high-quality, standardized images from diverse sources, which could greatly improve the accuracy and reliability of radiomic studies. By leveraging the strengths of both standardization and synthesis models, researchers may be able to unlock new insights into the relationship between CT images and disease outcomes.


\section{Method} \label{sec_method}
The structure of DiffusionCT is shown in Fig~\ref{fig:diff_model}, encompassing two major components: the image embedding component and a conditional DDPM in the latent space. The image embedding component employs an encoder-decoder network to translate input CT images to a low-dimensional latent representation. Subsequently, the conditional DDPM models the conditional probability distribution of the latent representation in order to synthesize a standard image. Importantly, DiffusionCT retains the original structure of the input image while effectively standardizing its texture.

DiffusionCT is trained sequentially in three steps. First, in the pre-processing step, the encoder-decoder network is trained with all CT images in the training set, irrespective of whether they are standard or non-standard or whether they are captured using GE or Siemens. This step aims to effectively encode images into a 1-D latent vector, which can reconstruct the original image with minimal information loss. Second, a latent conditional DDPM is trained with image pairs, consisting of a non-standard image and its corresponding standard image. This step enables the DDPM to model the conditional probability distribution of the latent representation, thus facilitating the synthesis of standard images. Finally, all the trained neural networks are combined to standardize new images.

\subsection{Image encoding and decoding}
\label{sec_model}
The image embedding component of DiffusionCT comprises a customized U-Net structured convolutional network, designed to learn a low-dimensional latent representation of input images. The encoder and decoder of the U-Net are asymmetric. 
The encoder uses a pre-trained ResNet-18 with four neural blocks. The first convolutional block consists of the first three ResNet-18 layers. The second block consists of the fourth and fifth layers of ResNet-18. The 3rd, 4th, and 5th blocks of the encoder consist of the corresponding 5th, 6th, and 7th layers of ResNet-18, respectively. The decoder encompasses a five-block convolutional network with up-sampling and several 1D convolutional layers in the last layers. Skip connection is not used within the 1D convolutional layers. 

This novel U-Net is trained with all available images in the training dataset, irrespective of whether they are standard or non-standard, in order to learn a global image encoding. The anatomic loss is adopted to facilitate the learning of structural information within the images. The trained U-Net encodes an input image into a latent low-dimensional representation, and the decoder accepts a latent representation to reconstruct the input image. This step is applicable for both intra-scanner and cross-vendor image standardization. The L2 loss is adopted for model training. 

\subsection{Conditional latent DDPM}

\begin{figure}[bt!]
\centering
\includegraphics[width=.5\textwidth]{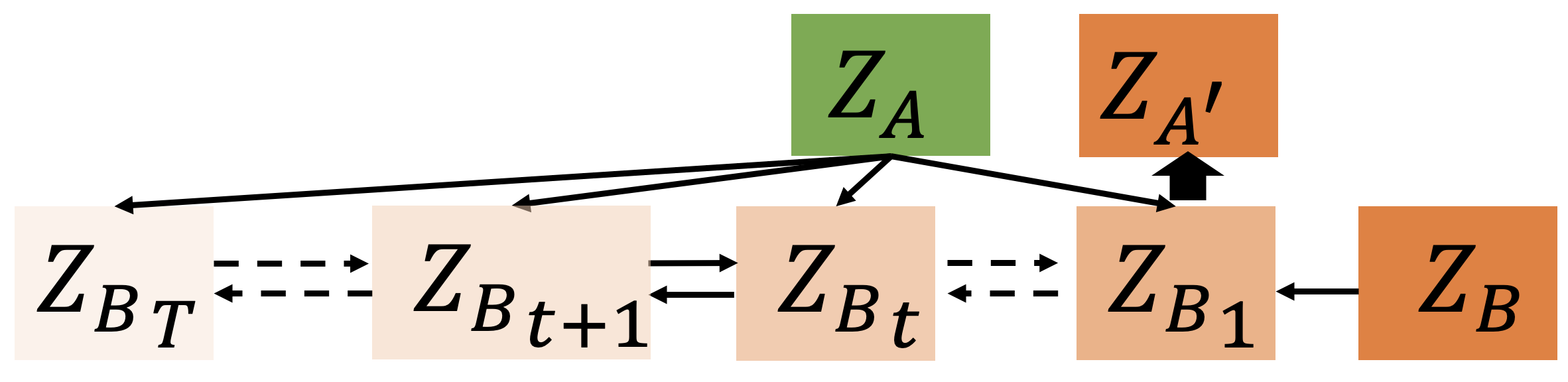}
\caption{\textbf{Conditional latent DDPM} for converting embedding $Z_A$ to $Z_{A'}$ in the B domain. } \label{fig:diff_lat}
\end{figure}

In the context of intra-scanner image standardization, paired image data are provided, consisting of a non-standard image $A$ and the corresponding standard image $B$. Using the previously described trained encoder, latent embeddings $Z_A$ and $Z_B$ are generated from non-standard ($A$) and standard ($B$) images, respectively. As $Z_A$ and $Z_B$ adhere to distinct distributions, a conditional latent DDPM is designed to map the non-standard latent distribution to the standard latent distribution. The encoder-decoder network remains unaltered during diffusion training. A well-trained conditional latent DDPM preserves anatomic details in $Z_A$ while mapping texture details from $Z_A$ to $Z_B$. 

Structure-wise, the conditional latent DDPM includes multiple small steps of diffusion in each training step. In every individual diffusion step, Gaussian noise $\eta$ is added to the latent embedding $Z_B$. All the corrupted $Z_B$  conditioned to $Z_A$ are used to train the conditional latent DDPM described in Figure~\ref{fig:diff_lat}. For a significant large $T$, where $T$ represents the total number of diffusion steps, $\prod_{t-1}^{T}(Z_{B_t}+\eta)$ converges to an isotropic Gaussian distribution. 

The network structure of the conditional latent DDPM is a U-Net, which is trained to predict the added noise $\eta$ from $\prod_{t-1}^{T}(Z_{B_t}+\eta)$. In addition to the standard diffusion loss function (see details at Ho et al~\cite{ho2020denoising}), an L1-loss between the reconstructed and the non-standard embeddings $\mathcal{L} = \mathbb{E}_{t\sim [1,T]} [|\eta_t  - p_\theta(Z_A,Z_{B_t})|]$ is used to update the diffusion model. After training, for each non-standard embedding $Z_A$, the model synthesizes a latent standardized embedding $Z_{A'}$.

\subsection{Model training}
To ensure effective training, we consider a two-step strategy, i.e., representation learning and latent diffusion training. In representation learning, we train the customized U-Net with all training images to learn the latent low-dimensional representation. Specifically, the network introduced in \ref{sec_model} is trained to learn the global data representation of all training images in the latent space. After the encoder and decoder are well trained, they remain fixed, and the latent diffusion model training starts. 
In the latent diffusion training process, we train the proposed conditional latent DDPM introduced in~\ref{sec_model} to map the latent representation of non-standardized images to the standard image domain. 

The trained encoder-decoder network and conditional latent DDPM are integrated for image standardization. A non-standard image $A$ is passed through the trained encoder to convert it into a latent representation $Z_A$. Then, $Z_A$ is passed through the trained conditional latent DDPM to generate $Z_{A'}$, which falls into the standard embedding domain. Finally, $Z_{A'}$ is passed through the trained decoder to synthesize image $A'$ in the standard image domain $B$.

\section{Experimental Results}
DiffusionCT was implemented using PyTorch on a Linux computer server equipped with eight Nvidia GTX 1080 GPU cards. The network weights were randomly initialized. The learning rate was set to $10^{-4}$ with the Adam optimizer. The encode-decode network underwent training for a duration of 20 epochs, followed by an additional 20 epochs dedicated to training the diffusion network. The model took about 20 hours to train from scratch. Once the model was trained, it took about 30 seconds to process and synthesize a DICOM CT image slice. The source code and related documents are available at \url{https://github.com/selimmd/DiffusionCT}.

We compared DiffusionCT with five recently developed CT image standardization models, including GANai~\cite{liang2019ganai}, STAN-CT~\cite{selim2020stan}, and, RadiomicGAN~\cite{selim2021radiomic}, CVH-CT~\cite{selim2021cross}, and UDA-CT~\cite{selim2021cross}, as well as the original DDPM and the encoder-decoder network.  
%
To evaluate the model performance, the results were measured using two metrics: the concordance correlation coefficient (CCC) and relative error (RE). These metrics allow for a quantitative evaluation of the effectiveness of the proposed method in achieving CT image standardization while preserving the original texture and structure of the images.

\subsection{Experimental Data} 
%

The training data consist of a total of 9,886 CT image slices from 14 lung cancer patients captured using two different kernels (Br40 and Bl64) and 1mm slice thickness using a Siemens CT Somatom Force scanner at the University of Kentucky Albert B. Chandler Hospital. 
The training data also contain additional 9,900 image slices from a Lungman chest phantom scan, with three synthetic tumors inserted. The phantom is scanned using two different kernels (Br40 and Bl64) and two different slice thicknesses of 1.5mm and 3mm using the same scanner. In total, 19,786 CT image slices were used to train DiffusionCT. 
To prepare the testing data, the identical Lungman chest phantom was used. The testing data comprised  126 CT image slices acquired using two different kernels (Br40 and Bl64) with a Siemens CT Somatom Force scanner. Notably, despite the commonality of the phantom used in obtaining both training and testing data sets, the acquisition of test data with a 5mm slice thickness results in the disjoint nature of the training and testing data. 
In this experiment, for demonstration purposes, Siemens Bl64 is considered the standard protocol, while Siemens Br40 was regarded as non-standard. 



\begin{figure*}[!bt]
\centering
\includegraphics[width=\textwidth]{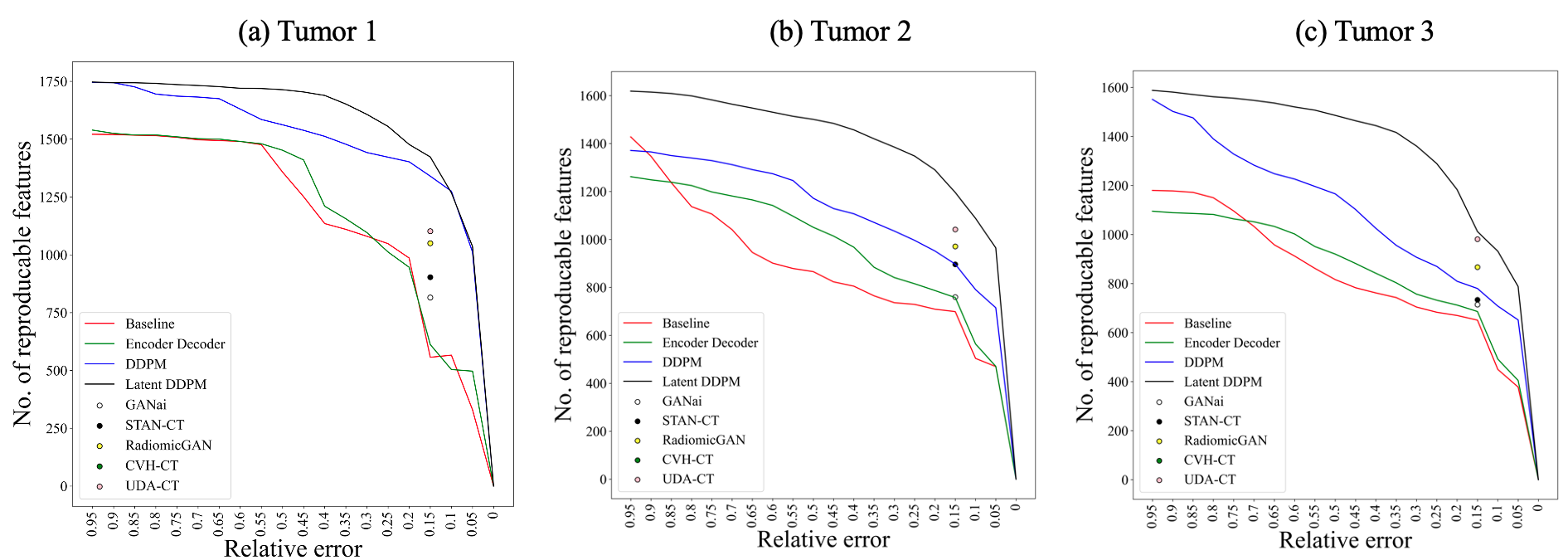}
\caption{\textbf{Total number of reproducible features after Siemens Br40 image synthesis.} Each point on the line represents the total number of reproducible features for the respective error threshold. The existing models' performances are denoted by the circle-shaped points only for the error threshold $RE<0.15$. } \label{fig:se_re}
\end{figure*}

\subsection{Evaluation Metric}
Model performance was evaluated based on lung tumors in the CT images. For each tumor, a total of 1,401 radiomic features, from six feature classes (GOH, GLCM, GLRLM, ID, IH, NID), were extracted using IBEX~\cite{zhang2015ibex}. 
Based on these radiomic features, we evaluated DiffusionCT and all the baseline models using two evaluation metrics, with one-to-one feature comparison and group-wise comparison.  

First, the relative error (RE), defined as the relative difference between a synthesized image and its corresponding standard image regarding a radiomic feature, was utilized to calculate the linear distance between the standard and the synthesized images regarding each individual radiomic feature. RE ranges from 0 to 1, and is the lower the better.  
\begin{equation}\label{eq:error_distance}
RE(s,t) = \frac{|f_t - f_s|}{f_t}
\end{equation}
\noindent where $f_t$ and $f_s$ are the radiomic feature values of the standard and synthesized image, respectively;  and $s$ and $t$ stand for the standard and the synthesized images, respectively.

Usually, a radiomic feature is considered to be reproducible if the synthesized image is more than 85\% similar to the corresponding standard image~\cite{zhao2016reproducibility,choe2019deep}. Mathematically, a radiomic feature is considered reproducible if and only if its relative error $RE(s,t) < 0.15$. 

Concordance Correlation Coefficient~\cite{lawrence1989concordance} (CCC) was employed to measure the level of similarity between two feature groups~\cite{choe2019deep}. Mathematically, CCC represents the correlation between the standard and the non-standard image features in the radiomic feature class $r$. CCC ranges from -1 to 1, and is the higher the better.  
\begin{equation}\label{eq:ccc}
    CCC(s,t, r) = \frac{2\rho_{s,t,r} \sigma _s \sigma_t}{{\sigma _s}^2 +{\sigma_t}^2 + {( \mu _s - \mu_t )}^2}
\end{equation}
\noindent where $s$ and $t$ stand for the standard and the synthesized images, respectively; $\mu_s$ and $\sigma_s$ (or $\mu_t$ and $\sigma_t$) are the mean and standard deviation of the radiomic features belonging to the same feature class $R$ in a synthesized (or standard) image, respectively; and $\rho_{s,t,r}$ is the Pearson correlation coefficient between $s$ and $t$ regarding a feature class $r$.

\subsection{Results and Discussion}

\begin{table*}[!bt]
 	\centering
 	\caption { The CCC values of the images synthesized using different image standardization models. Each column represents the mean$\pm$std CCC values of a specific radiomic feature group. }
 	\begin{tabular}{ c  c  c  c c  c  c    }
 		\hline
Feature Class	&	GOH			&	GLCM			&	GLRLM			&	ID			&	IH			&	NID			\\ \hline

Baseline	&	0.90	$\pm$	0.05	&	0.20	$\pm$	0.13	&	0.59	$\pm$	0.13	&	0.33	$\pm$	0.16	&	0.35	$\pm$	0.12	&	0.28	$\pm$	0.15	\\ 
GANai	&	0.95	$\pm$	0.05	&	0.50	$\pm$	0.08	&	0.63	$\pm$	0.12	&	0.59	$\pm$	0.03	&	0.44	$\pm$	0.08	&	0.65	$\pm$	0.10	\\ 
STAN-CT	&	0.95	$\pm$	0.05	&	0.70	$\pm$	0.10	&	0.72	$\pm$	0.15	&	0.75	$\pm$	0.16	&	0.61	$\pm$	0.11	&	0.71	$\pm$	0.05	\\ 
RadiomicGAN	&	1.00	$\pm$	0.00	&	0.80	$\pm$	0.12	&	0.75	$\pm$	0.11	&	0.82	$\pm$	0.08	&	0.72	$\pm$	0.09	&	0.73	$\pm$	0.12	\\ \hline				
Encoder-Decoder	&	1.00	$\pm$	0.00	&	0.38	$\pm$	0.19	&	0.61	$\pm$	0.15	&	0.52	$\pm$	0.11	&	0.39	$\pm$	0.25	&	0.33	$\pm$	0.09	\\ \hline			
DDPM	&	1.00	$\pm$	0.00	&	0.81	$\pm$	0.23	&	0.80	$\pm$	0.18	&	0.85	$\pm$	0.15	&	0.77	$\pm$	0.12	&	0.82	$\pm$	0.13	\\ 
DiffusionCT	&	1.00	$\pm$	0.00	&	0.85	$\pm$	0.14	&	0.79	$\pm$	0.21	&	0.89	$\pm$	0.28	&	0.41	$\pm$	0.05	&	0.86	$\pm$	0.18	\\ \hline	

 	\end{tabular} \label{table:a_dataset}
\end{table*}

In Figure~\ref{fig:se_re}, each point on a line represents the total number of radiomic features on the y-axis whose respective relative error is equal to or smaller than the value specified  on the x-axis.
The red line represents the direct comparison of the input images and the corresponding standard images without using any algorithms. The green, blue, and black lines represent the performance of the encoder-decoder network, DDPM, and DiffusionCT model, respectively. In the literature, the compared models' performances were reported based on a 15\% relative error. In figure~\ref{fig:se_re}, the model performance on $RE\leq 0.15$ showed that DiffusionCT preserved 64\% and DDPM preserved 58\% more radiomic features than the baseline, comparing to GANai at 20\%, STAN-CT at 32\%, and RadiomicGAN at 51\%. 

Table~\ref{table:a_dataset} shows the CCC scores of six  classes of radiomic features. The performance of the baseline was measured using the input images. In four out of six feature classes, DiffusionCT achieved $CCC >0.85$, clearly outperforming all the compared models. Nevertheless,  DDPM outperformed DiffusionCT and other compared models in two other feature groups. Notably, GLCM and GLRLM together occupy almost 50\% of the total number of radiomic features, and both the DDPM and our DiffusionCT achieved significant performance gains. Also, DDPM had the highest variation on GLCM, indicating conditional DDPM could be more suitable for the image standardization task

\begin{figure*}[!tb]
\centering
\includegraphics[width=.8\textwidth]{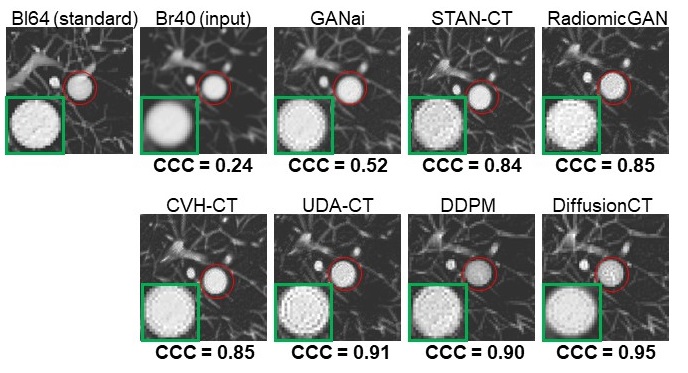}
\caption{CT images synthesized using all compared models in the display window of [-800, 600] HU. The leftmost image is the standard image, and the right bottom is the result of DiffusionCT. Each image contains the same ROI with a tumor marked in a red circle and magnified in the green box. CCC scores of GLCM are displayed at the bottom.} \label{fig:example}
\end{figure*}



Figure~\ref{fig:example} visualizes the results of all compared models on a sample tumor. The input tumor image is observably different from the standard image regarding visual appearances as well as radiomic features. The DiffusionCT-generated image has the highest CCC values regarding GLCM in reference to the standard image and is visually more similar to the standard image than the ones generated by GAN-based models and the vanilla DDPM.

\section{Conclusion}
Image standardization reduces texture feature variations and improves the reliability of radiomic features of CT imaging. The existing CT image standardization models were mainly developed based on GAN. This article accesses the application DDPM approach for the CT image standardization task. Both image space and latent space have been investigated in relation to DDPM. The experimental results indicate that DDPM-based models are significantly better than GAN-based models. The DDPM has comparable performance in image space and latent space. Owing to its relatively compact size, DiffusionCT is best suited for creating more abstract embeddings in the target domain. 

\section*{Acknowledgements} This research is supported by NIH NCI (grant no. 1R21CA231911) and Kentucky Lung Cancer Research (grant no. KLCR-3048113817).

\makeatletter
\renewcommand{\@biblabel}[1]{\hfill #1.}
\makeatother

\bibliographystyle{unsrt}
\bibliography{bibfile} 

\begin{thebibliography}{10}

\bibitem{collins2017letter}
Jannette Collins.
\newblock Letter from the editor: Lung cancer screening facts.
\newblock In {\em Seminars in roentgenology}, volume~52, pages 121--122, 2017.

\bibitem{de2014benefits}
Harry~J De~Koning, Rafael Meza, Sylvia~K Plevritis, Kevin Ten~Haaf, Vidit~N
  Munshi, Jihyoun Jeon, Saadet~Ayca Erdogan, Chung~Yin Kong, Summer~S Han,
  Joost Van~Rosmalen, et~al.
\newblock Benefits and harms of computed tomography lung cancer screening
  strategies: a comparative modeling study for the us preventive services task
  force.
\newblock {\em Annals of internal medicine}, 160(5):311--320, 2014.

\bibitem{ravanelli2013texture}
Marco Ravanelli, Davide Farina, Mauro Morassi, Elisa Roca, Giuseppe Cavalleri,
  Gianfranco Tassi, and Roberto Maroldi.
\newblock Texture analysis of advanced non-small cell lung cancer (nsclc) on
  contrast-enhanced computed tomography: prediction of the response to the
  first-line chemotherapy.
\newblock {\em European radiology}, 23:3450--3455, 2013.

\bibitem{ardila2019end}
Diego Ardila, Atilla~P Kiraly, Sujeeth Bharadwaj, Bokyung Choi, Joshua~J
  Reicher, Lily Peng, Daniel Tse, Mozziyar Etemadi, Wenxing Ye, Greg Corrado,
  et~al.
\newblock End-to-end lung cancer screening with three-dimensional deep learning
  on low-dose chest computed tomography.
\newblock {\em Nature medicine}, 25(6):954--961, 2019.

\bibitem{song2017using}
QingZeng Song, Lei Zhao, XingKe Luo, and XueChen Dou.
\newblock Using deep learning for classification of lung nodules on computed
  tomography images.
\newblock {\em Journal of healthcare engineering}, 2017, 2017.

\bibitem{lu2020deep}
Michael~T Lu, Vineet~K Raghu, Thomas Mayrhofer, Hugo~JWL Aerts, and Udo
  Hoffmann.
\newblock Deep learning using chest radiographs to identify high-risk smokers
  for lung cancer screening computed tomography: development and validation of
  a prediction model.
\newblock {\em Annals of Internal Medicine}, 173(9):704--713, 2020.

\bibitem{berenguer2018radiomics}
Roberto Berenguer, Mar{\'\i}a del~Rosario Pastor-Juan, Jes{\'u}s
  Canales-V{\'a}zquez, Miguel Castro-Garc{\'\i}a, Mar{\'\i}a~Victoria Villas,
  Francisco~Mansilla Legorburo, and Sebasti{\`a} Sabater.
\newblock Radiomics of ct features may be nonreproducible and redundant:
  Influence of ct acquisition parameters.
\newblock {\em Radiology}, page 172361, 2018.

\bibitem{hunter2013high}
Luke~A Hunter, Shane Krafft, Francesco Stingo, Haesun Choi, Mary~K Martel,
  Stephen~F Kry, and Laurence~E Court.
\newblock High quality machine-robust image features: Identification in
  nonsmall cell lung cancer computed tomography images.
\newblock {\em Medical physics}, 40(12), 2013.

\bibitem{paul2012relationships}
Jijo Paul, B~Krauss, R~Banckwitz, et~al.
\newblock Relationships of clinical protocols and reconstruction kernels with
  image quality and radiation dose in a 128-slice ct scanner: study with an
  anthropomorphic and water phantom.
\newblock {\em European journal of radiology}, 81(5):e699--e703, 2012.

\bibitem{gierada2010effects}
David~S Gierada, Andrew~J Bierhals, Cliff~K Choong, Seth~T Bartel, Jon~H
  Ritter, Nitin~A Das, Cheng Hong, Thomas~K Pilgram, Kyongtae~T Bae, Bruce~R
  Whiting, et~al.
\newblock Effects of ct section thickness and reconstruction kernel on
  emphysema quantification: relationship to the magnitude of the ct emphysema
  index.
\newblock {\em Academic radiology}, 17(2):146--156, 2010.

\bibitem{ours_aamp}
G~Liang, J~Zhang, M~Brooks, J~Howard, and J~Chen.
\newblock radiomic features of lung cancer and their dependency on ct image
  acquisition parameters.
\newblock {\em Medical Physics}, 44(6):3024, 2017.

\bibitem{cohen2012radiosity}
Michael~F Cohen and John~R Wallace.
\newblock {\em Radiosity and realistic image synthesis}.
\newblock Elsevier, 2012.

\bibitem{selim2020stan}
Md~Selim, Jie Zhang, Baowei Fei, Guo-Qiang Zhang, and Jin Chen.
\newblock Stan-ct: Standardizing ct image using generative adversarial network.
\newblock In {\em AMIA Annual Symposium Proceedings}, volume 2020. American
  Medical Informatics Association, 2020.

\bibitem{selim2021cross}
Md~Selim, Jie Zhang, Baowei Fei, and et. al.
\newblock Cross-vendor ct image data harmonization using cvh-ct.
\newblock In {\em AMIA Annual Symposium Proceedings}, volume 2021, page 1099.
  American Medical Informatics Association, 2021.

\bibitem{selim2021radiomic}
Md~Selim, Jie Zhang, and et. al.
\newblock {CT} image harmonization for enhancing radiomics studies.
\newblock In {\em 2021 IEEE International Conference on Bioinformatics and
  Biomedicine (BIBM)}, pages 1057--1062, 2021.

\bibitem{ho2020denoising}
Jonathan Ho, Ajay Jain, and Pieter Abbeel.
\newblock Denoising diffusion probabilistic models.
\newblock {\em Advances in Neural Information Processing Systems},
  33:6840--6851, 2020.

\bibitem{saharia2022image}
Chitwan Saharia, Jonathan Ho, William Chan, Tim Salimans, David~J Fleet, and
  Mohammad Norouzi.
\newblock Image super-resolution via iterative refinement.
\newblock {\em IEEE Transactions on Pattern Analysis and Machine Intelligence},
  2022.

\bibitem{saharia2022palette}
Chitwan Saharia, William Chan, Huiwen Chang, Chris Lee, Jonathan Ho, Tim
  Salimans, David Fleet, and Mohammad Norouzi.
\newblock Palette: Image-to-image diffusion models.
\newblock In {\em ACM SIGGRAPH 2022 Conference Proceedings}, pages 1--10, 2022.

\bibitem{yip2016applications}
Stephen~SF Yip and Hugo~JWL Aerts.
\newblock Applications and limitations of radiomics.
\newblock {\em Physics in Medicine \& Biology}, 61(13):R150, 2016.

\bibitem{yang2011quantifying}
Xiangyu Yang and Michael~V Knopp.
\newblock Quantifying tumor vascular heterogeneity with dynamic
  contrast-enhanced magnetic resonance imaging: a review.
\newblock {\em BioMed Research International}, 2011, 2011.

\bibitem{basu2011evolving}
Sandip Basu, Thomas~C Kwee, Robert Gatenby, Babak Saboury, Drew~A Torigian, and
  Abass Alavi.
\newblock Evolving role of molecular imaging with pet in detecting and
  characterizing heterogeneity of cancer tissue at the primary and metastatic
  sites, a plausible explanation for failed attempts to cure malignant
  disorders, 2011.

\bibitem{liang2019ganai}
Gongbo Liang, Sajjad Fouladvand, Jie Zhang, Michael~A Brooks, Nathan Jacobs,
  and Jin Chen.
\newblock Ganai: Standardizing ct images using generative adversarial network
  with alternative improvement.
\newblock In {\em 2019 IEEE International Conference on Healthcare Informatics
  (ICHI)}, pages 1--11. IEEE, 2019.

\bibitem{pix2pix}
Phillip Isola, Jun-Yan Zhu, Tinghui Zhou, and Alexei~A. Efros.
\newblock Image-to-image translation with conditional adversarial networks.
\newblock In {\em Computer Vision and Pattern Recognition (CVPR)}, 2017.

\bibitem{gonzalez2012digital}
Rafael~C Gonzalez and Richard~E Woods.
\newblock {\em Digital image processing}.
\newblock Upper Saddle River, NJ: Prentice Hall, 2012.

\bibitem{CycleGAN2017}
Jun-Yan Zhu, Taesung Park, Phillip Isola, and Alexei~A Efros.
\newblock Unpaired image-to-image translation using cycle-consistent
  adversarial networks.
\newblock In {\em 2017 IEEE International Conference on Computer Vision
  (ICCV)}, pages 2242--2251, 2017.

\bibitem{selim2022udact}
Md~Selim, Jie Zhang, and et. al.
\newblock {UDA-CT}: A general framework for ct image standardization.
\newblock In {\em 2022 IEEE International Conference on Bioinformatics and
  Biomedicine (BIBM)}, 2022.

\bibitem{zhang2015ibex}
Lifei Zhang, David~V Fried, Xenia~J Fave, and et. al.
\newblock Ibex: an open infrastructure software platform to facilitate
  collaborative work in radiomics.
\newblock {\em Medical physics}, 42(3):1341--1353, 2015.

\bibitem{zhao2016reproducibility}
Binsheng Zhao, Yongqiang Tan, Wei-Yann Tsai, Jing Qi, Chuanmiao Xie, Lin Lu,
  and Lawrence~H Schwartz.
\newblock Reproducibility of radiomics for deciphering tumor phenotype with
  imaging.
\newblock {\em Scientific reports}, 6(1):1--7, 2016.

\bibitem{choe2019deep}
Jooae Choe, Sang~Min Lee, Kyung-Hyun Do, Gaeun Lee, June-Goo Lee, Sang~Min Lee,
  and Joon~Beom Seo.
\newblock Deep learning--based image conversion of ct reconstruction kernels
  improves radiomics reproducibility for pulmonary nodules or masses.
\newblock {\em Radiology}, 292(2):365--373, 2019.

\bibitem{lawrence1989concordance}
I~Lawrence and Kuei Lin.
\newblock A concordance correlation coefficient to evaluate reproducibility.
\newblock {\em Biometrics}, pages 255--268, 1989.

\end{thebibliography}

\end{document}